\documentclass[12pt,onecolumn]{article}
\usepackage[dvips]{graphicx}
\usepackage{amsfonts}
\usepackage{indentfirst}
\usepackage{epsf}

\title{Quantum-information theoretic properties of nuclei
and trapped Bose gases}
\date{June, 2005}

\author{Ch.~C.~Moustakidis\footnote{\texttt{email:moustaki@auth.gr}},
K.~Ch.~Chatzisavvas\footnote{\texttt{email:kchatz@auth.gr}},
C.~P.~Panos\footnote{\texttt{email:chpanos@auth.gr}}
\\
 {\it  Physics Department,}\\
 {\it Aristotle University of Thessaloniki,}\\
 {\it  54124 Thessaloniki, Greece}
 }

\begin{document}

\maketitle

\begin{abstract}
Fermionic (atomic nuclei) and bosonic (correlated atoms in a trap)
systems are studied from an information-theoretic point of view.
Shannon and Onicescu information measures are calculated for the
above systems comparing correlated and uncorrelated cases as
functions of the strength of short range correlations. One-body
and two-body density and momentum distributions are employed. Thus
the effect of short-range correlations on the information content
is evaluated. The magnitude of distinguishability of the
correlated and uncorrelated densities is also discussed employing
suitable measures of distance of states i.e. the well known
Kullback-Leibler relative entropy and the recently proposed
Jensen-Shannon divergence entropy. It is seen that the same
information-theoretic properties hold for quantum many-body
systems obeying different statistics (fermions and bosons).
\end{abstract}

\section{Introduction}

Information-theoretic methods are used in recent years for the
study of quantum mechanical systems.
\cite{Ohya}$^{-}$\cite{Chatzisavvas} The quantity of interest is
Shannon's information entropy for a probability distribution
$p(x)$
\begin{equation}\label{eq:equ1}
    S=-\int p(x)\,\ln{p(x)}\,dx
\end{equation}
where $\int p(x)\,dx=1$.

An important step is the discovery of an entropic uncertainty
relation (EUR),\cite{Bialynicki} which for a three-dimensional
system has the form
\begin{equation}\label{eq:equ2}
    S=S_r+S_k\geq 3\,(1+\ln{\pi})\simeq 6.434
\end{equation}
where $S_r$ is the information entropy in position-space of the
density distribution $\rho(\textbf{r})$ of a quantum system
\begin{equation}\label{eq:equ3}
    S_r=-\int \rho(\textbf{r})\,\ln{\rho(\textbf{r})}\,d\textbf{r}
\end{equation}
and $S_k$ is the information entropy in momentum-space of the
corresponding momentum distribution $n(\textbf{k})$
\begin{equation}\label{eq:equ4}
    S_k=-\int n(\textbf{k})\,\ln{n(\textbf{k})}\,d\textbf{k}
\end{equation}

The density distributions $\rho(\textbf{r})$ and $n(\textbf{k})$
are normalized to one. Inequality (\ref{eq:equ2}), for the
information entropy sum in conjugate spaces, is a joint measure of
uncertainty of a quantum mechanical distribution, since a highly
localized $\rho(\textbf{r})$ is associated with a diffuse
$n(\textbf{k})$, leading to low $S_r$ and high $S_k$ and
vice-versa. Expression (\ref{eq:equ2}) is an
information-theoretical relation stronger than Heisenberg's. $S$
is measured in bits if the base of the logarithm is 2 and nats
(natural units of information) if the logarithm is natural.

In previous work we proposed a universal property of $S$ for the
density distributions of nuclei, electrons in atoms and valence
electrons in atomic clusters.\cite{Massen2} This property has the
form
\begin{equation}\label{eq:equ5}
    S=a+b \ln{N}
\end{equation}
where $N$ is the number of particles of the system and the
parameters $a, b$ depend on the system under consideration. It is
noted that recently we have obtained the same form for systems of
correlated bosons in a trap.\cite{Massen1} This concept was also
found to be useful in a different context. Using the formalism in
phase-space of Ghosh, Berkowitz and Parr,\cite{Ghosh} we found
that the larger the information entropy the better the quality of
the nuclear density distribution.\cite{Lalazissis}

In previous work we employed one-body density distributions in the
definition of $S$. In the present paper we introduce two-body
density distributions $\rho(\textbf{r}_1,\textbf{r}_2)$ and the
corresponding two-body momentum distributions
$n(\textbf{k}_1,\textbf{k}_2)$. Our aim is to investigate the
properties of $S$ at the two-body level for correlated densities.
The correlated nucleon systems or the trapped Bose gas, in a good
approximation, are studied using the lowest order
approximation.\cite{Fabrocini,Moustakidis2} Short-range
correlations (SRC) are taken into account employing the Jastrow
correlation function.\cite{Jastrow} Thus it is of interest to
examine how $S_2$ is affected qualitatively and quantitatively by
the same form of correlations in comparison with $S_1$, in view of
the fact that the quantities $\rho(\textbf{r}_1,\textbf{r}_2)$ and
$n(\textbf{k}_1,\textbf{k}_2)$ carry more direct information for
correlations than the quantities $\rho(\textbf{r})$ and
$n(\textbf{k})$ which are only indirectly affected by
correlations. The above procedure is repeated for an alternative
measure of information i.e. Onicescu's information energy
$E$.\cite{Onicescu} So far, only the mathematical aspects of this
concept have been developed, while the physical aspects have been
neglected.

A well known measure of distance of two discrete probability
distributions $p_i^{(1)}, p_i^{(2)}$ is the Kullback-Leibler
relative entropy \cite{Kullback}
\begin{equation}\label{eq:equ6}
    K(p_i^{(1)},p_i^{(2)})=\sum_i p_i^{(1)}\,\ln{\frac{p_i^{(1)}}{p_i^{(2)}}}
\end{equation}
which for continuous probability distributions $\rho^{(1)},
\rho^{(2)}$ is defined as
\begin{equation}\label{eq:equ7}
    K=\int \rho^{(1)}(x)\,\ln{\frac{\rho^{(1)}(x)}{\rho^{(2)}(x)}}\,dx
\end{equation}
which can be easily extended for 3-dimensional systems.

Our aim is to calculate the relative entropy (distance) between
$p^{(1)}$ (correlated) and $p^{(2)}$ (uncorrelated) densities both
at the one- and the two-body levels in order to assess the
influence of SRC (through the correlation parameter $y$) on the
distance $K$. It is noted that this is done for both systems under
consideration: nuclei and trapped Bose gases. An alternative
definition of distance of two probability distributions was
introduced by Rao and Lin,\cite{Rao,Lin} i.e. a symmetrized
version of $K$, the Jensen-Shannon divergence $J$ \cite{Majtey}
\begin{equation}\label{eq:equ8}
    J(p^{(1)},p^{(2)})=H\left(\frac{p^{(1)}+p^{(2)}}{2}\right)-\frac{1}{2}H\left(p^{(1)}\right)
    -\frac{1}{2}H\left(p^{(2)}\right)
\end{equation}
where $H(p)=-\sum_i p_i \ln{p_i}$ stands for Shannon's entropy. We
expect for strong SRC the amount of distinguishability of the
correlated from the uncorrelated distributions is larger than the
corresponding one with small SRC. We may also see the effect of
SRC on the number of trials $L$ needed to distinguish $p^{(1)}$
and $p^{(2)}$ (in the sense described in \cite{Majtey}).

In addition to the above considerations, we connect $S_r$ and
$S_k$ with fundamental quantities i.e. the root mean square radius
and kinetic energy respectively. We also argue on the effect of
SRC on EUR and we propose a universal relation for $S$, by
extending our formalism from the one- and two-body level to the
$N$-body level, which holds exactly for uncorrelated densities in
trapped Bose gas, almost exactly for uncorrelated densities in
nuclei (due to the additional exchange term compared to Bose gas)
and it is conjectured to hold approximately for correlated
densities both in nuclei and Bose gases.

The plan of the present paper is the following. In Sec.
\ref{sec:2} we review the formulas of Kullback-Leibler relative
entropy entropy $K$ and Jensen-Shannon divergence $J$, while in
Sec. \ref{sec:3} Onicescu's information energy $E$ is described.
In Sec. \ref{sec:4} we present the formalism of density
distributions used in present work and their applications to
Shannon's and Onicescu's entropies. In Sec. \ref{sec:5} we
introduce SRC in nuclei. In Sec. \ref{sec:6} we apply the formulas
of $K$ and $J$ in correlated distributions. In Sec. \ref{sec:8} we
present our numerical results and discussion. Finally, Sec.
\ref{sec:9} contains our main conclusions.

\section{Kullback-Leibler relative entropy and Jensen-Shannon
divergence} \label{sec:2}

The Kullback-Leibler relative information entropy $K$ for
continuous distributions $\rho_i^{(1)}$ and $\rho_i^{(2)}$ is
defined by relation (\ref{eq:equ7}). It measures the difference of
$\rho_i^{(1)}$ form the reference (or apriori) distribution
$\rho_i^{(2)}$. It satisfies: $K\geq 0$ for any distributions
$\rho_i^{(1)}$ and $\rho_i^{(2)}$. It is a measure which
quantifies the distinguishability (or distance) of $\rho_i^{(1)}$
from $\rho_i^{(2)}$, employing a well-known concept in standard
information theory. In other words it describes how close
$\rho_i^{(1)}$ is to $\rho_i^{(2)}$ by carrying out observations
or coin tossing, namely $L$ trials (in the sense described in
\cite{Majtey}). We expect for strong SRC the amount of
distinguishability of the correlated $\rho_i^{(1)}$ and the
uncorrelated distributions $\rho_i^{(2)}$ is larger than the
corresponding one with small SRC.

However, the distance $K$ does not satisfy the triangle inequality
and in addition is i) not symmetric ii) unbounded and iii) not
always well defined.\cite{Majtey} To avoid these difficulties Rao
and Lin \cite{Rao,Lin} introduced a symmetrized version of $K$
(recently discused in \cite{Majtey}), the Jensen-Shannon
divergence $J$ defined by relation (\ref{eq:equ8}). $J$ is minimum
for $\rho^{(1)}=\rho^{(2)}$ and maximum when $\rho^{(1)}$ and
$\rho^{(2)}$ are two distinct distributions, when $J=\ln{2}$. In
our case  $J$ can be easily generalized for continuous density
distributions. For $J$ minimum the two states represented by
$\rho^{(1)}$ and $\rho^{(2)}$ are completely indistinguishable,
while for $J$ maximum they are completely distinguishable. It is
expected that for strong SRC the amount of distinguishability can
be further examined by using Wooter's criterion.\cite{Majtey} Two
probability distributions $\rho^{(1)}$ and $\rho^{(2)}$ are
distinguishable after $L$ trials $(L\rightarrow \infty)$ if and
only if $\left( J(\rho^{(1)},\rho^{(2)})
\right)^{\frac{1}{2}}>\frac{1}{\sqrt{2L}}$.

The present work is a first step to examine the problem of
comparison of probability distributions (for nuclei and bosonic
systems) which is an area well developed in statistics, known as
information geometry.\cite{Rao}

\section{Onicescu's information energy}\label{sec:3}

Onicescu tried to define a finer measure of dispersion
distributions than that of Shannon's information
entropy.\cite{Onicescu} Thus, he introduced the concept of
information energy $E$. For a discrete probability distribution
$(p_1,p_2,\ldots,p_k)$ the information energy $E$ is defined by
\begin{equation}\label{eq:equ9}
    E=\sum_i^k p_i^2
\end{equation}
which is extended for a continuous density distribution $\rho(x)$
as
\begin{equation}\label{eq:equ10}
    E=\int \rho^2(x)\,dx
\end{equation}
The meaning of (\ref{eq:equ10}) can be seen by the following
simple argument: For a Gaussian distribution of mean value $\mu$,
standard deviation $\sigma$ and normalized density
\begin{equation}\label{eq:equ11}
  \rho(x)=\frac{1}{\sqrt{2\pi}\sigma}\, \textrm{exp}
  \left[-\frac{(x-\mu)^2}{2\sigma^2} \right]
\end{equation}
relation (\ref{eq:equ10}) gives
\begin{equation}\label{eq:equ12}
  E=\frac{1}{2\pi\sigma^2} \int_{-\infty}^{\infty}
   \textrm{exp}
  \left[-\frac{(x-\mu)^2}{\sigma^2}
  \right]\,dx=\frac{1}{2\sigma\sqrt{\pi}}
\end{equation}
$E$ is maximum if one of the $p_i$'s equals 1 and all the others
are equal to zero i.e. $E_{max}=1$, while $E$ is minimum when
$p_1=p_2=\ldots=p_k=\frac{1}{k}$, hence $E_{min}=\frac{1}{k}$
(total disorder). The fact that $E$ becomes minimum for equal
probabilities (total disorder), by analogy with thermodynamics, it
has been called information energy, although it does not have the
dimension of energy.\cite{Lepadatu}

It is seen from (\ref{eq:equ12}) that the greater the information
energy, the more concentrated is the probability distribution,
while the information content decreases. $E$ and information
content are reciprocal, hence one can define the quantity
\begin{equation}\label{eq:equ13}
  O=\frac{1}{E}
\end{equation}
as a measure of the information content of a quantum system
corresponding to Onicescu's information energy.

Relation (\ref{eq:equ10}) is extended for a 3-dimensional
spherically symmetric density distribution $\rho(\textbf{r})$
\begin{eqnarray}\label{eq:equ14}
  E_r=\int \rho^2(\textbf{r})\,d\textbf{r} \nonumber \\
  E_k=\int n^2(\textbf{k})\,d\textbf{k}
\end{eqnarray}
in position and momentum space respectively, where $n(\textbf{k})$
is the corresponding density distribution in momentum space.

$E_r$ has dimension of inverse volume, while $E_k$ of volume. Thus
the product $E_r E_k$ is dimensionless and can serve as a measure
of concentration (or information content) of a quantum system. It
is also seen from (\ref{eq:equ12}),(\ref{eq:equ13}) that $E$
increases as $\sigma$ decreases (or concentration increases) and
the information (or uncertainty) decreases. Thus $O$ and $E$ are
reciprocal. In order to be able to compare $O$ with Shannon's
entropy $S$, we redifine $O$ as
\begin{equation}\label{eq:equ15}
  O=\frac{1}{E_r E_k}
\end{equation}
as a measure of the information content of a quantum system in
both position and momentum spaces, inspired by Onicescu's
definition.

\section{Density Matrices and Information entropies}\label{sec:4}

Let $\Psi({\bf r}_1,{\bf r}_2,\cdots,{\bf r}_A)$ be the wave
function that describes the nuclei or the trapped Bose gases  and
depends on 3A coordinates as well as on spin and isospin (in
nuclei). The one-body density matrix is defined in \cite{Lowdin}

\begin{equation}
\rho({\bf r}_1,{\bf r}_1')=\int \Psi^{*}({\bf r}_1,{\bf
r}_2,\cdots,{\bf r}_A) \Psi({\bf r}_1',{\bf r}_2,\cdots,{\bf r}_A)
d{\bf r}_2 \cdots d{\bf r}_A \label{OBDM-1}
\end{equation}
while the two-body density matrix by
\begin{equation}
\rho({\bf r}_1,{\bf r}_2;{\bf r}_1',{\bf r}_2')= \int
\Psi^{*}({\bf r}_1,{\bf r}_2,\cdots,{\bf r}_A) \Psi({\bf
r}_1',{\bf r}_2',\cdots,{\bf r}_A) d{\bf r}_3 \cdots d{\bf r}_A
\label{TBDM-1}
\end{equation}
The above density matrices are related by
\begin{equation}
\rho({\bf r}_1,{\bf r}_1')=\frac{1}{A-1} \int \rho({\bf r}_1,{\bf
r}_2;{\bf r}_1',{\bf r}_2)  d{\bf r}_2 \label{O-T-1}
\end{equation}
where the integration is carried out over the radius vectors ${\bf
r}_2,\cdots,{\bf r}_A$ and summation over spin (or isospin)
variables is implied. The corresponding definitions in momentum
space are similar. The two-body density distribution $\rho({\bf
r}_1,{\bf r}_2)$  which is a key quantity in the present work, is
defined as the diagonal part of the two-body density matrix
\begin{equation}
\rho({\bf r}_1,{\bf r}_2)= \rho({\bf r}_1,{\bf r}_2;{\bf
r}_1',{\bf r}_2')\mid_{{\bf r}_1'={\bf r}_1, {\bf r}_2'={\bf r}_2}
\label{TBDD-1}
\end{equation}
and expresses the joint probability of finding two nucleons or two
atoms at the positions ${\bf r}_1$ and ${\bf r}_2$, respectively.
The density distribution is given by the diagonal part of the
one-body density matrix, that is
\begin{equation}
\rho({\bf r}_1)=\rho({\bf r}_1,{\bf r}_1')|_{{\bf r}_1={\bf r}_1'}
\label{DD-1}
\end{equation}
or by the equivalent integral
\begin{equation}
\rho({\bf r}_1)=\frac{1}{A-1}  \int \rho({\bf r}_1,{\bf r}_2)
 d{\bf r}_2
 \label{DD-2}
 \end{equation}

The two-body momentum distribution $n({\bf k}_1,{\bf k}_2)$ is
given by a particular Fourier transform of the $ \rho({\bf
r}_1,{\bf r}_2;{\bf r}_1',{\bf r}_2')$, that is
\begin{equation}
n({\bf k}_1,{\bf k}_2)=\frac{1}{(2\pi)^6} \int \rho({\bf r}_1,{\bf
r}_2;{\bf r}_1',{\bf r}_2') \exp[i{\bf k}_1({\bf r}_1-{\bf r}_1')]
\exp[i{\bf k}_2({\bf r}_2-{\bf r}_2')] d {\bf r}_1 d {\bf r}_1' d
{\bf r}_2 d {\bf r}_2' \label{TBMD-1}
\end{equation}

In the independent particle model, where the nucleons are
considered to move independently in nuclei, the $\Psi({\bf
r}_1,{\bf r}_2,\cdots,{\bf r}_A)$ is a Slater determinant. In this
case it is easy to show that the two-body density matrix is given
by the relation
\begin{eqnarray}
\rho_{SD}({\bf r}_1,{\bf r}_2;{\bf r}_1',{\bf r}_2')&=&
\sum_{i,j}\phi_i({\bf r}_1)\phi_i({\bf r}_1') \phi_j({\bf
r}_2)\phi_j({\bf r}_2')- \sum_{i,j}\phi_i({\bf r}_1)\phi_j({\bf
r}_1')
\phi_j({\bf r}_2)\phi_i({\bf r}_2')\nonumber \\
&=& \rho_{SD}({\bf r}_1,{\bf r}_1')\rho_{SD}({\bf r}_2,{\bf
r}_2')- \rho_{SD}({\bf r}_1,{\bf r}_2') \rho_{SD}({\bf r}_2,{\bf
r}_1') \label{TBDM-SD}
\end{eqnarray}
where $\phi_i({\bf r})$ is the single-particle wave function
normalized to one and
\[
   \rho_{SD}({\bf r}_1,{\bf
   r}_1')=\sum_{i}\phi_i({\bf r}_1)\phi_i({\bf r}_1')
\]

In Bose gases the many-body ground-state wave function $\Psi({\bf
r}_1,{\bf r}_2,\cdots,{\bf r}_A)$ is a product of $A$ identical
single-particle ground-state wave functions i.e.
\begin{equation}
\Psi({\bf r}_1,{\bf r}_2,\cdots,{\bf r}_A)= \phi_{0}({\bf
r}_1)\phi_{0}({\bf r}_2)\cdots \phi_{0}({\bf r}_A) \label{WF-BG}
\end{equation}
where $\phi_{0}({\bf r}_1)$ is the normalized to one ground-state
single-particle wave function describing  bosonic atoms. The
two-body density matrix in a Bose gas, is given by the relation
\begin{equation}
\rho_{0}({\bf r}_1,{\bf r}_2;{\bf r}_1',{\bf r}_2')= \rho_{0}({\bf
r}_1,{\bf r}_1')\rho_{0}({\bf r}_2,{\bf r}_2') \label{TBDM-BG}
\end{equation}
where
\begin{equation}
\rho_{0}({\bf r}_1,{\bf r}_1')=\phi_{0}({\bf r}_1) \phi_{0}({\bf
r}_1')
\end{equation}
We consider that the atoms of the Bose gases are confined in an
isotropic HO well, where $\phi_0({\bf r})=(1/(\pi b^2))^{3/4}
\exp[-r^2/(2b^2)]$.

As the mean field approach fails to incorporate the interparticle
correlation which is necessary for the description of the
correlated nuclei or trapped Bose gases, we introduce the
repulsive interactions through the Jastrow correlation function
$f({\bf r}_1-{\bf r}_2)$ \cite{Jastrow}. The correlated nucleon
systems or the Bose gases, in a good approximation, can be studied
using the lowest order approximation,\cite{Fabrocini,Moustakidis2}
where the correlated two-body density matrices in nuclei and Bose
gases have the following forms respectively
\begin{equation}
\rho({\bf r}_1,{\bf r}_2;{\bf r}_1',{\bf r}_2')=N \rho_{SD}({\bf
r}_1,{\bf r}_2;{\bf r}_1',{\bf r}_2') f({\bf r}_1-{\bf r}_2)
f({\bf r}_1'-{\bf r}_2') \label{TBDM-2}
\end{equation}
\begin{equation}
\rho({\bf r}_1,{\bf r}_2;{\bf r}_1',{\bf r}_2')=N \rho_{0}({\bf
r}_1,{\bf r}_2;{\bf r}_1',{\bf r}_2') f({\bf r}_1-{\bf r}_2)
f({\bf r}_1'-{\bf r}_2') \label{TBDM-2}
\end{equation}

In the present work, in the case of nuclei and trapped Bose gas,
the normalization factor $N$, is calculated by the normalization
condition
\begin{equation}\label{eq:norm1}
  \int\rho(\textbf{r}_1,\textbf{r}_2)\,d\textbf{r}_1\,d\textbf{r}_2=1
\end{equation}
The same holds for $n(\textbf{k}_1,\textbf{k}_2)$
\begin{equation}\label{eq:norm2}
  \int
  n(\textbf{k}_1,\textbf{k}_2)\,d\textbf{k}_1\,d\textbf{k}_2=1
\end{equation}

The Jastrow correlation function $f(\textbf{r}_1-\textbf{r}_2)$
both in the case of nuclei and trapped Bose gas is taken to be of
the form
\begin{equation}\label{eq:jastrow}
  f(\textbf{r}_1-\textbf{r}_2)=1-
  \textrm{exp}[-y\,\frac{(\textbf{r}_1-\textbf{r}_2)^2}{b^2}]
\end{equation}

The uncorrelated case corresponds to $y\rightarrow \infty$, while
SRC increase as $y$ decreases. The above ansatz has the advantage
that it leads to analytical forms for the
$\rho(\textbf{r}_1,\textbf{r}_2)$, $n(\textbf{k}_1,\textbf{k}_2)$,
$\rho(\textbf{r})$ and $n(\textbf{k})$.

The one-body Shannon information entropy both in position- and
momentum- space are defined in (\ref{eq:equ3}) and
(\ref{eq:equ4}), where the total sum is
\begin{equation}
S_1=S_{1r}+S_{1k} \label{S1-1}
\end{equation}

The two-body Shannon information entropy both in position- and
momentum- space and in total are defined respectively
\cite{Amovilli,Cover}
\begin{equation}
S_{2r}=-\int \rho({\bf r}_1,{\bf r}_2) \ln \rho({\bf r}_1,{\bf
r}_2) d {\bf r}_1 d{\bf r}_2 \label{S2r-1}
\end{equation}
\begin{equation}
S_{2k}=-\int n({\bf k}_1,{\bf k}_2) \ln n({\bf k}_1,{\bf k}_2)
d{\bf k}_1 d{\bf k}_2 \label{S2k-1}
\end{equation}
\begin{equation}
S_2=S_{2r}+S_{2k} \label{S2-1}
\end{equation}

The one-body Onicescu information entropy is already defined in
(\ref{eq:equ14}) and (\ref{eq:equ15}), where the generalization to
the two-body information entropy is straightforward and is given
by
\begin{equation}
O_2= \frac{1} {E_{2r} E_{2k}} \label{O1-2}
\end{equation}
where
\begin{eqnarray}
E_{2r}&=&\int \rho^2({\bf r}_1,{\bf r}_2) d{\bf r}_1 d{\bf r}_2 \nonumber \\
E_{2k}&=&\int n^2({\bf k}_1,{\bf k}_2) d{\bf k}_1 d {\bf k}_2
\label{E2r-2k}
\end{eqnarray}

It is easy to prove that in the case of the uncorrelated trapped
Bose gas
\begin{equation}
S_2=2 S_1 \label{S1-S2}
\end{equation}
and
\begin{equation}
O_2=O_1^2 \label{O1-O2}
\end{equation}

It is worth noting that the above relations hold only
approximately in finite nuclei (see Table \ref{tab:table1}), due
to the additional exchange term, originating from the antisymmetry
of the nuclear wave function. There is an exception in the case of
$^4$He, where it holds exactly due to the absence of the exchange
term.

\section{Introduction of SRC in nuclei}\label{sec:5}
We consider that the single particle wave functions, which
describe the nucleons is harmonic oscillator type. In order to
incorporate the nucleon-nucleon (or atom-atom) correlations, as we
mention in the previous section, we apply the lowest order
approximation. In this case the two-body density distribution, for
$^4$He, takes the following form
\begin{equation}
\rho^{^4He}({\bf r}_1,{\bf r}_2)=\rho^{^4He}_{SD}({\bf r}_1,{\bf
r}_2)+ \rho^{^4He}_{cor}({\bf r}_1,{\bf r}_2) \label{2den-he-1}
\end{equation}
The first term of the right-hand side of Eq. (\ref{2den-he-1})
which represents the uncorrelated part of the two-body sensity
distribution, has the form
\begin{equation}
\rho^{^4He}_{SD}({\bf r}_1,{\bf r}_2)= \frac{1}{\pi^3 b^6}
\exp[-r_{1b}^2]\exp[-r_{2b}^2] \label{TBDM-unc}
\end{equation}
and the second term which represents the correlated part of the
two-body density distribution, is written
\begin{eqnarray}
\rho^{^4He}_{cor}({\bf r}_1,{\bf r}_2)&=& \frac{1}{\pi^3 b^6}
\exp[-r_{1b}^2]\exp[-r_{2b}^2] \nonumber \\
& &\times \left( N \left(1-\exp[-y({\bf r}_{1b}-{\bf r}_{2b})^2 ]
\right)^2-1 \right)  \label{TBDM-cor}
\end{eqnarray}
where $\textbf{r}_{b}={\bf r}/b$.

In the above expression $b$ is the width of the HO potential and
$N$ is the normalization constant which ensures that $\int
\rho^{^4He}_{cor}({\bf r}_1,{\bf r}_2) d{\bf r}_1 d{\bf r}_2=1$
and has the form
\begin{equation}
N=\left(1-\frac{2}{(1+2y)^{3/2}}+\frac{1}{(1+4y)^{3/2}}\right)^{-1}
\label{norm}
\end{equation}
The density distribution can be written also in the form
\begin{equation}
\rho^{^4He}(r)=\rho_{SD}^{^4He}(r)+\rho_{cor}^{^4He}(r)
\label{den-he4-1}
\end{equation}
The two-body momentum distribution is given also by the formula
\begin{equation}
n^{^4He}({\bf k}_1,{\bf k}_2)=n^{^4He}_{SD}({\bf k}_1,{\bf k}_2)+
n^{^4 He}_{cor}({\bf k}_1,{\bf k}_2) \label{2mom-he-1}
\end{equation}
where, as in the case of two-body density distribution, the
uncorrelated part has the form
\begin{equation}
n^{^4He}_{SD}({\bf k}_1,{\bf k}_2)= \frac{b^6}{\pi^3}
\exp[-k_{1b}^2] \exp[-k_{2b}^2] \label{TBMD-unc}
\end{equation}
and the correlated part is written as
\begin{eqnarray}
n^{^4He}_{cor}({\bf k}_1,{\bf k}_2)&=& \frac{b^6}{\pi^3}
\exp[-k_{1b}^2] \exp[-k_{2b}^2]  \\
& & \times \left( N(1-\frac{1}{(1+4y)^{3/2}}
\exp[-\frac{y}{1+4y}({\bf k}_{1b}-{\bf k}_{2b})^2 ])^2-1 \right)
\nonumber \label{TBMD-cor}
\end{eqnarray}
where $\textbf{k}_{b}=\textbf{k}\,b$. \\
The momentum distribution is given also by the relation
\begin{equation}
n^{^4 He}(k)=n_{SD}^{^4 He}(k)+n_{cor}^{^4 He}(k)
\label{1mom-he4-1}
\end{equation}

In the present work, we extend  our calculations in nuclei heavier
than $^4\textrm{He}$ ($^{12}$C, $^{16}$O and $^{40}$Ca) based on
the fact that the high-momentum tails of $n(k)$ are almost the
same for all nuclei with $A\geq 4$.\cite{Moustakidis,Massen6}
Inspired by previous work \cite{Stringari,Gaidarov} we suggest a
practical method to calculate the one- and two-body density and
momentum distributions for nuclei heavier than $^4$He. The
theoretical scheme of the method combines the mean-field
predictions of the two-body density distributions and two-body
momentum distributions of various nuclei with their correlated
part of $^4$He. Specifically, in our treatment we consider the
following forms
\begin{equation}
\rho^{A}({\bf r}_1,{\bf r}_2)=\rho^{A}_{SD}({\bf r}_1,{\bf r}_2)+
\rho^{^4\textrm{He}}_{cor}({\bf r}_1,{\bf r}_2) \label{2den-nuc-1}
\end{equation}
\begin{equation}
n^{A}({\bf k}_1,{\bf k}_2)=n^{A}_{SD}({\bf k}_1,{\bf k}_2)+
n^{^4\textrm{He}}_{cor}({\bf k}_1,{\bf k}_2) \label{2mom-nuc-1}
\end{equation}

From the above expressions it is obvious  that the uncorrelated
part of the $\rho(\textbf{r}_1,\textbf{r}_2)$ and
$n(\textbf{k}_1,\textbf{k}_2)$ originate from the independent
particle model for every nucleus separately, where the correlated
part in each nucleus is that coming from the nucleus $^4$He. The
$\rho(\textbf{r})$ and $n(\textbf{k})$ have a similar form.

It should be emphasized that in the uncorrelated case the
additional information which is contained in
$\rho(\textbf{r}_1,\textbf{r}_2)$ and
$n(\textbf{k}_1,\textbf{k}_2)$ in nuclei, compared to the trapped
Bose gas is the statistical correlations which come from the
antisymmetry character of the many-body wave function of nuclei.
Moreover, in the correlated case the
$\rho(\textbf{r}_1,\textbf{r}_2)$ and
$n(\textbf{k}_1,\textbf{k}_2)$ contain additional information
which originate from the character of the nuleon-nucleon
interaction, making our model more realistic and the description
more complete. It is of interest to study how the correlations
(both statistical and dynamical) affect quantitatively and
qualitatively the various kinds of information entropy.

\section{Application of the Formalism of Relative Entropy and Jensen-Shannon divergence
for Correlated Densities}\label{sec:6}

The relative entropy is a measure of distinguishability or
distance of two states. It is defined, generalizing
(\ref{eq:equ7}), by
\begin{equation}\label{eq:equ16}
    K=\int \psi^2(\textbf{r})
    \ln{\frac{\psi^2(\textbf{r})}{\phi^2(\textbf{r})}}\,d\textbf{r}
\end{equation}
In our case $\psi(\textbf{r})$ is the correlated case and
$\phi(\textbf{r})$ the uncorrelated one. Thus
\begin{equation}\label{eq:equ17}
    K_{1r}=\int \rho(\textbf{r})\,\ln{\frac{\rho(\textbf{r})}{\rho'(\textbf{r})}}
    \,d\textbf{r}
\end{equation}
where $\rho(\textbf{r})$ is the correlated one-body density and
$\rho'(\textbf{r})$ is the uncorrelated one-body density.

A corresponding formula holds in momentum-space
\begin{equation}\label{eq:equ18}
     K_{1k}=\int
     n(\textbf{k})\,\ln{\frac{n(\textbf{k})}{n'(\textbf{k})}}
     \,d\textbf{k}
\end{equation}
where $n(\textbf{k})$ is the correlated one-body density and
$n'(\textbf{k})$ is the uncorrelated one.

For the two-body case we have
\begin{equation}\label{eq:equ19}
    K_{2r}=\int \rho(\textbf{r}_1,\textbf{r}_2)\,
    \ln{\frac{\rho(\textbf{r}_1,\textbf{r}_2)}{\rho'(\textbf{r}_1,\textbf{r}_2)}}\,
    d\textbf{r}_1 d\textbf{r}_2
\end{equation}
where $\rho(\textbf{r}_1,\textbf{r}_2)$ is the correlated two-body
density in position-space and $\rho'(\textbf{r}_1,\textbf{r}_2)$
is the uncorrelated one.

The generalization to momentum- space is straightforward
\begin{equation}\label{eq:equ20}
    K_{2k}=\int n(\textbf{k}_1,\textbf{k}_2)
    \ln{\frac{n(\textbf{k}_1,\textbf{k}_2)}{n'(\textbf{k}_1,\textbf{k}_2)}}\,
    d\textbf{k}_1 d\textbf{k}_2
\end{equation}
where $n(\textbf{k}_1,\textbf{k}_2)$ is the correlated two-body
density in momentum-space and $n'(\textbf{k}_1,\textbf{k}_2)$ is
the uncorrelated one.

For the Jensen-Shannon divergence $J$ we may write formulas for
$J_1$ (one-body) and $J_2$ (two-body), employing definition
(\ref{eq:equ8}) and putting the corresponding correlated
$\rho^{(1)}$ and uncorrelated $\rho^{(2)}$ distributions in
position- and momentum- spaces. We calculate $K$ and $J$ in
position- and momentum- spaces, for nuclei and bosons.

\section{Numerical results and discussion}\label{sec:8}

For the sake of symmetry and simplicity we put the width of the HO
potential $b=1$. Actually for $b=1$ in the case of uncorrelated
case it is easy to see that $S_{1r}=S_{1k}$ and also
$S_{2r}=S_{2k}$ (the same holds for Onicescu entropy), while when
$b\neq 1$ there is a shift of the values of $S_{1r}$ and $S_{1k}$
by an additive factor $\ln{b^3}$. However, the value of $b$ does
not affect directly the total information entropy $S$ (and also
$O$). $S$ and $O$ are just functions of the correlation parameter
$y$.

\begin{table}[ht]\label{tab:table1}
\begin{center}
\begin{tabular}{lllll}
\hline Nucleus &$S_1$ &$S_2$& $O_1$ & $\sqrt{O_2}$ \\ \hline
  $^4$He   & 6.43418& 12.86836& 248.05& 248.05 \\
$^{12}$C & 7.50858& 15.00784& 922.60& 921.15 \\
$^{16}$O & 7.60692& 15.20890& 1057.25& 1055.77 \\
$^{40}$Ca& 8.43472& 16.88498& 2685.72& 2711.75 \\
\hline
\end{tabular}
\caption{The values of the Shannon and Onicescu information
entropy (both one and two body) for various nuclei $s$-$p$ and
$s$-$d$ shell nuclei.}
\end{center}
\end{table}

\begin{figure}[h]
 \centering
 \includegraphics[height=5.0cm,width=3.5cm]{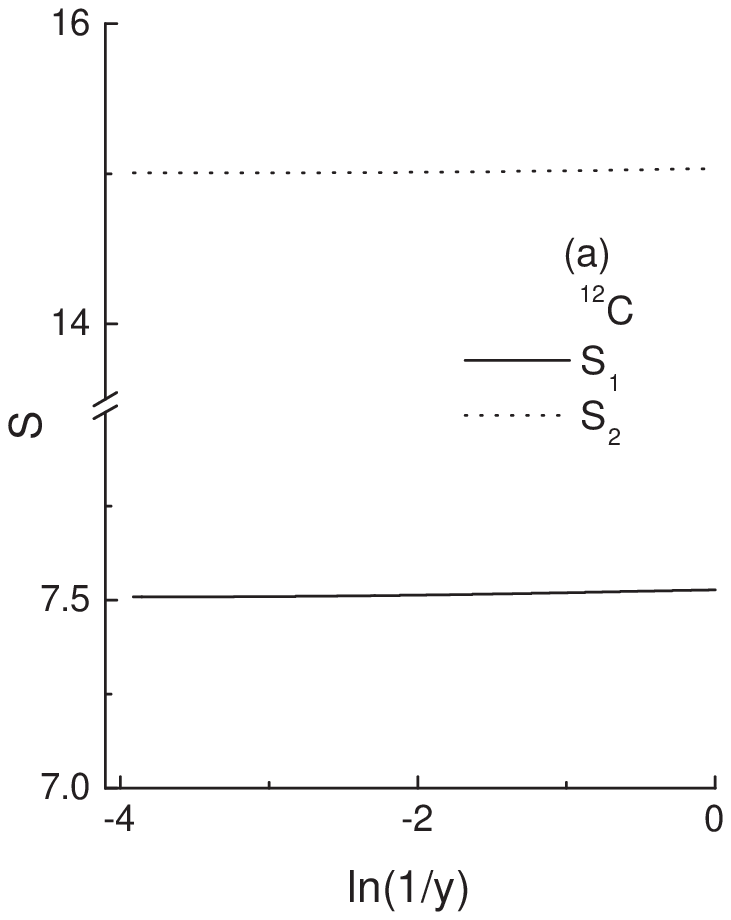}
 \hspace{0.5cm}
 \includegraphics[height=5.0cm,width=3.5cm]{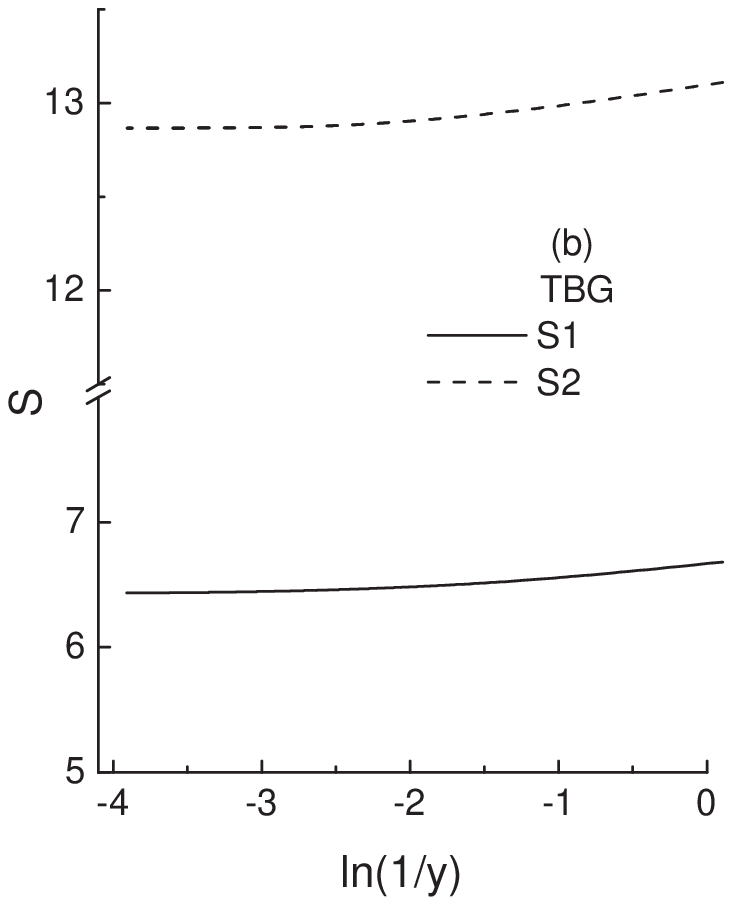}
 \caption{\label{fig:fig1} The Shannon information entropy one-body $S_1$ and two-body
$S_2$ (a) in nuclei $^{12}$C and (b) in a trapped Bose gas (TBG).}
\end{figure}

In Fig. \ref{fig:fig1} we present the Shannon information entropy
$S_1$ using relation (\ref{S1-1}) and $S_2$ using relation
(\ref{S2-1}) in nuclei $(^{12}\textrm{C})$ and trapped Bose gas as
functions of the correlation parameter $\ln{(\frac{1}{y})}$. It is
seen that $S_1$ and $S_2$ increase almost linearly with the
strength of SRC i.e. $\ln{(\frac{1}{y})}$ in both systems. The
relations $S_2=2 S_1$ and $O_2=O_1^2$ hold exactly for the
uncorrelated densities in trapped Bose gas, while the above
relations are almost exact for the uncorrelated densities in
nuclei and in the case of correlated densities both in nuclei and
trapped Bose gas. A similar behavior is seen for all nuclei
considered in the present work ($^{4}$He, $^{16}$O, $^{40}$Ca).

Values of $S_1$, $S_2$, $O_1$, $O_2$ for various nuclei in the
uncorrelated case, are shown in Table 1. The relations
(\ref{S1-S2}) and (\ref{O1-O2}) are satisfied exactly only in the
case of $^4$He. However, for the other nuclei, due to the
additional exchange term in the nuclear wave function, the
relations (\ref{S1-S2}) and (\ref{O1-O2}) hold only approximately
(the differences are of order $0.03\%-0.09\%$ for $S$ and
$0.14\%-0.96\%$ for $O$).

\begin{figure}[h]
 \centering
 \includegraphics[height=5.0cm,width=3.5cm]{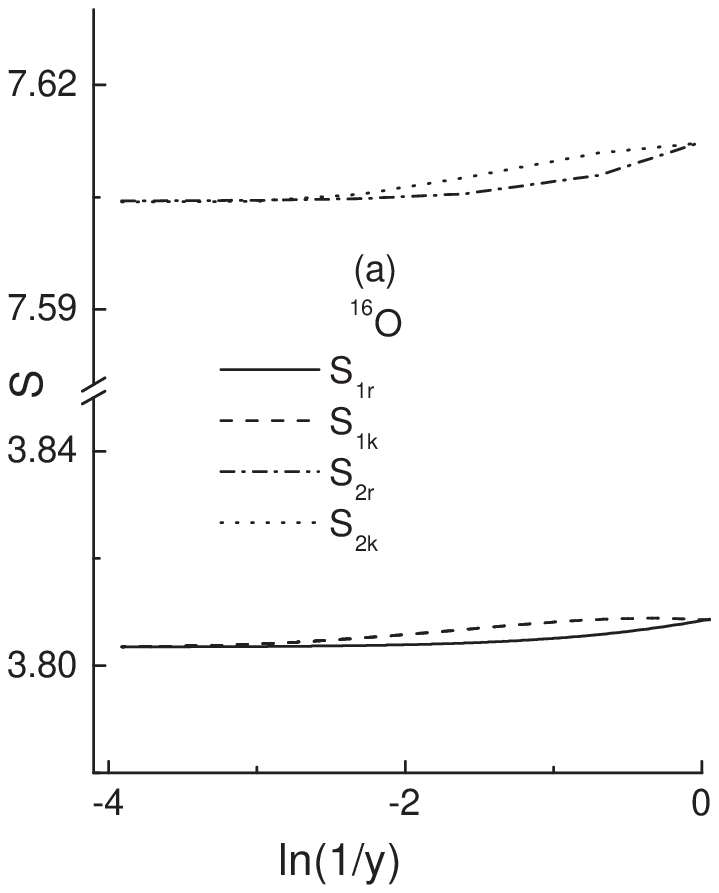}
 \hspace{0.5cm}
 \includegraphics[height=5.0cm,width=3.5cm]{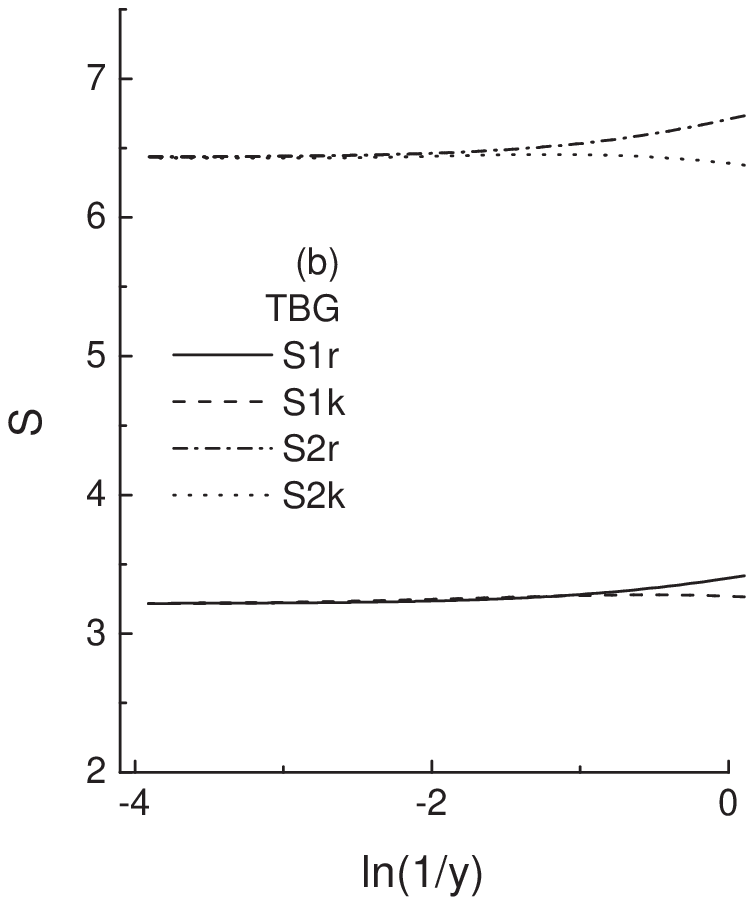}
 \caption{\label{fig:fig2} The Shannon information entropy (one- and two-body) both in
coordinate- and momentum-space (a) in nuclei $^{16}$O and (b) in a
trapped Bose gas (TBG).}
\end{figure}

In Fig. \ref{fig:fig2} we present the decomposition of $S$ in
coordinate and momentum spaces, for the sake of comparison i.e.
$S_{1r}$, $S_{1k}$, $S_{2r}$, $S_{2k}$ for $^{16}\textrm{O}$ and
trapped Bose gas employing (\ref{eq:equ3}), (\ref{eq:equ4}),
(\ref{S2r-1}), (\ref{S2k-1}). The most striking feature concluded
from the above Figures is the similar behavior between $S_{1r}$
and $S_{2r}$ and also $S_{1k}$ and $S_{2k}$ respectively.

\begin{figure}
 \centering
 \includegraphics[height=5.0cm,width=3.5cm]{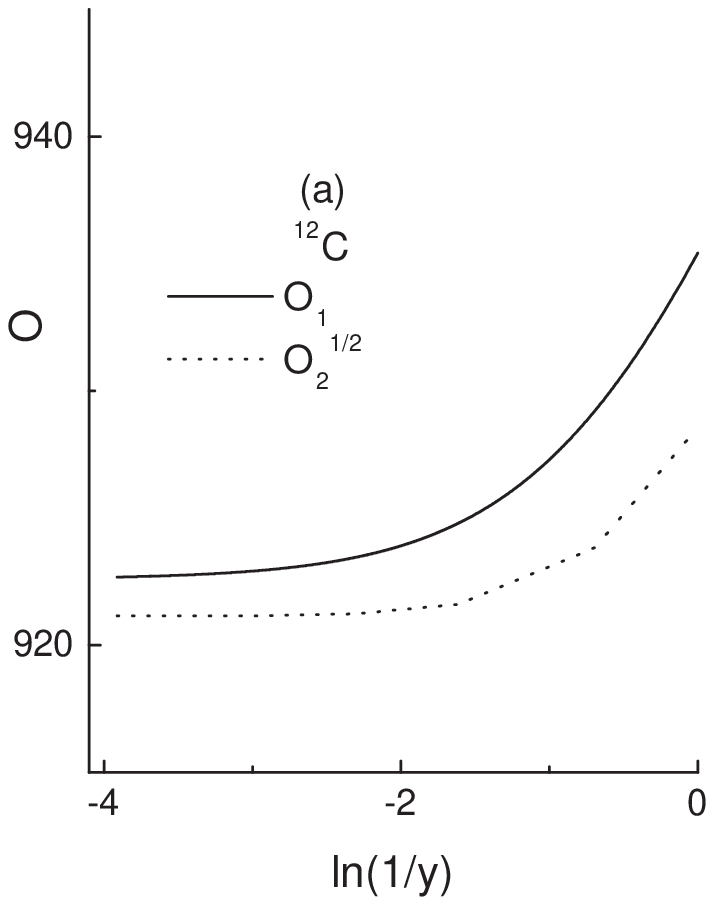}
 \hspace{0.5cm}
 \includegraphics[height=5.0cm,width=3.5cm]{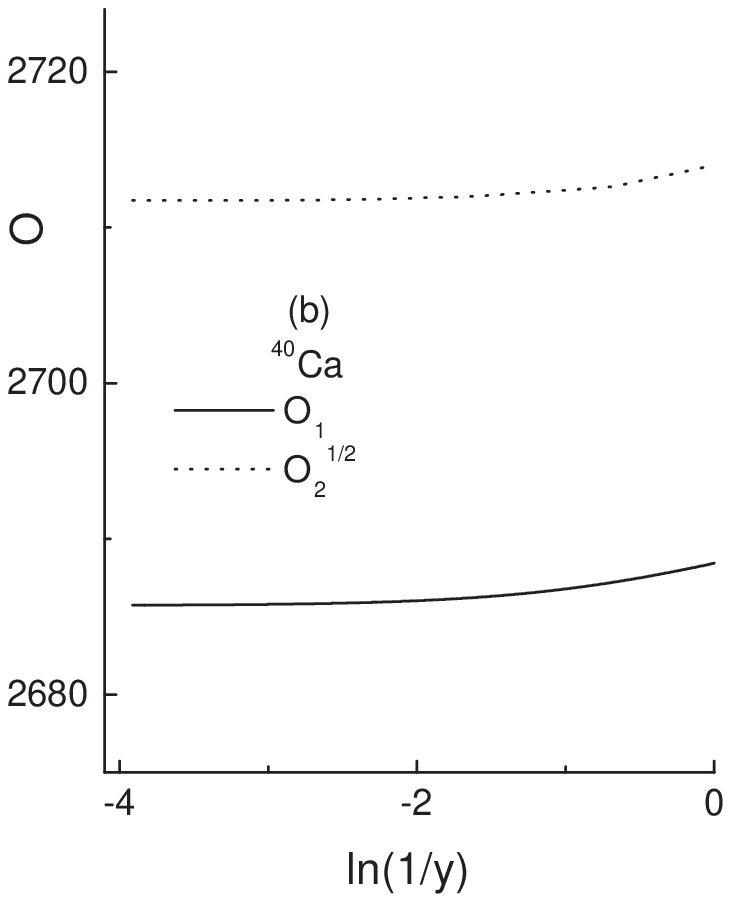}
 \hspace{0.5cm}
 \includegraphics[height=5.0cm,width=3.5cm]{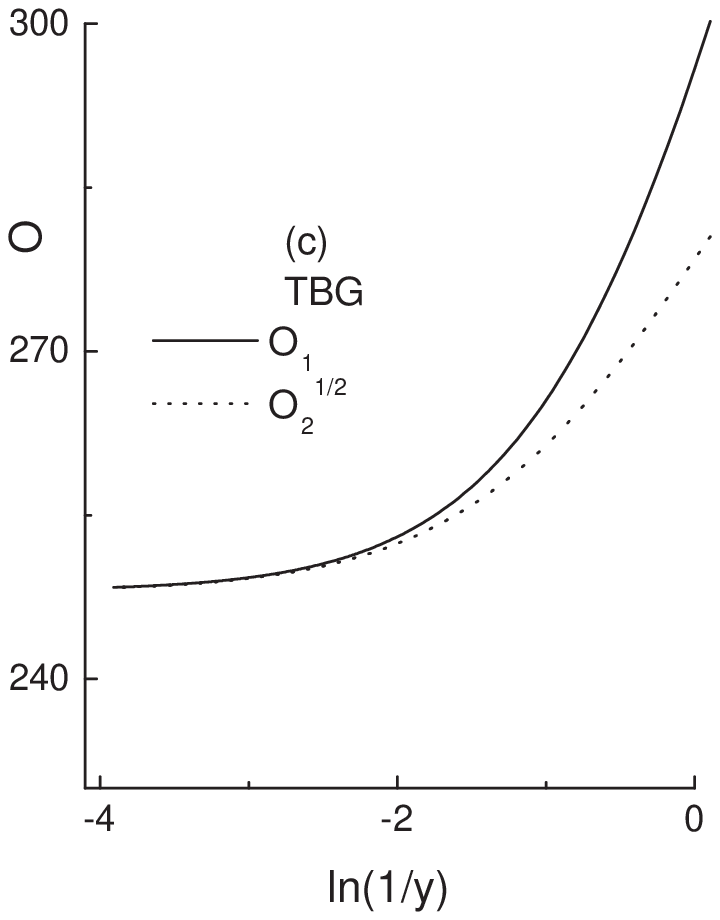}
 \caption{\label{fig:fig3} The Onicescu information entropy (both one- and two-body) (a)
in $^{12}$C, (b) in $^{40}$Ca and (c) in a trapped Bose gas
(TBG).}
\end{figure}

In Fig. \ref{fig:fig3} we plot the Onicescu information entropy
both one-body $(O_1)$ and two-body $(O_2)$ for nuclei
$(^{12}\textrm{C}, ^{40}\textrm{Ca})$ and trapped Bose gas
(relations (\ref{eq:equ15}), (\ref{O1-2})). We conclude by noting
once again the strong similarities of the behavior between one-
and two-body Onicescu entropy.

\begin{figure}[h]
 \centering
 \includegraphics[height=5.0cm,width=3.5cm]{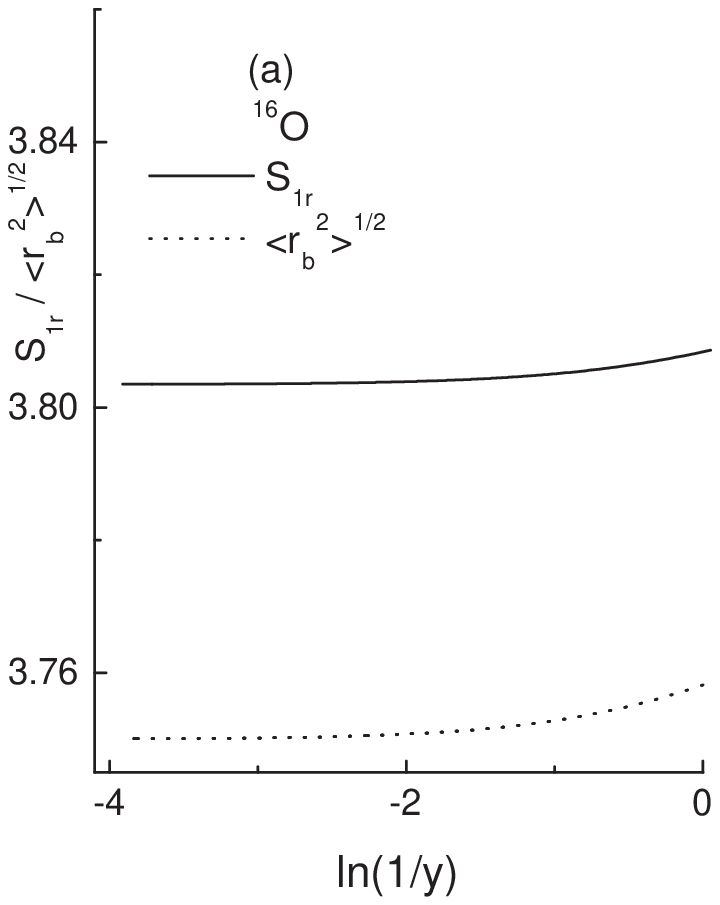}
 \hspace{0.5cm}
 \includegraphics[height=5.0cm,width=3.5cm]{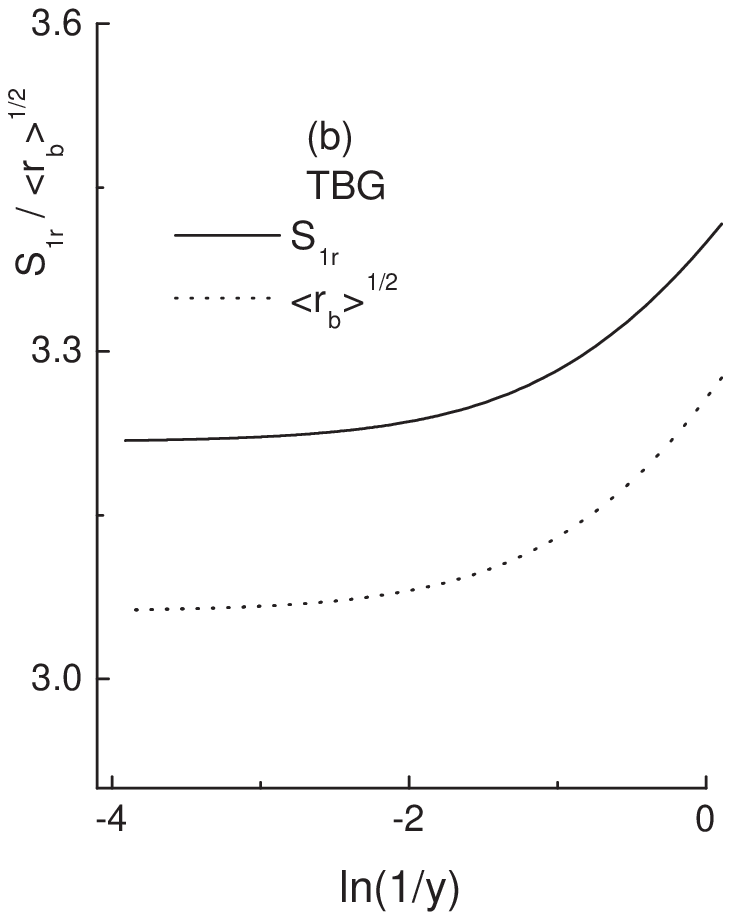}
 \caption{\label{fig:fig4} The mean-square radius and the Shannon information entropy
$S_{1r}$ as a function of the correlation parameter
$\ln{(\frac{1}{y})}$, (a) in nuclei $^{16}$O and (b) in a trapped
Bose gas (TBG).}
\end{figure}

\begin{figure}[h]
 \centering
 \includegraphics[height=5.0cm,width=3.5cm]{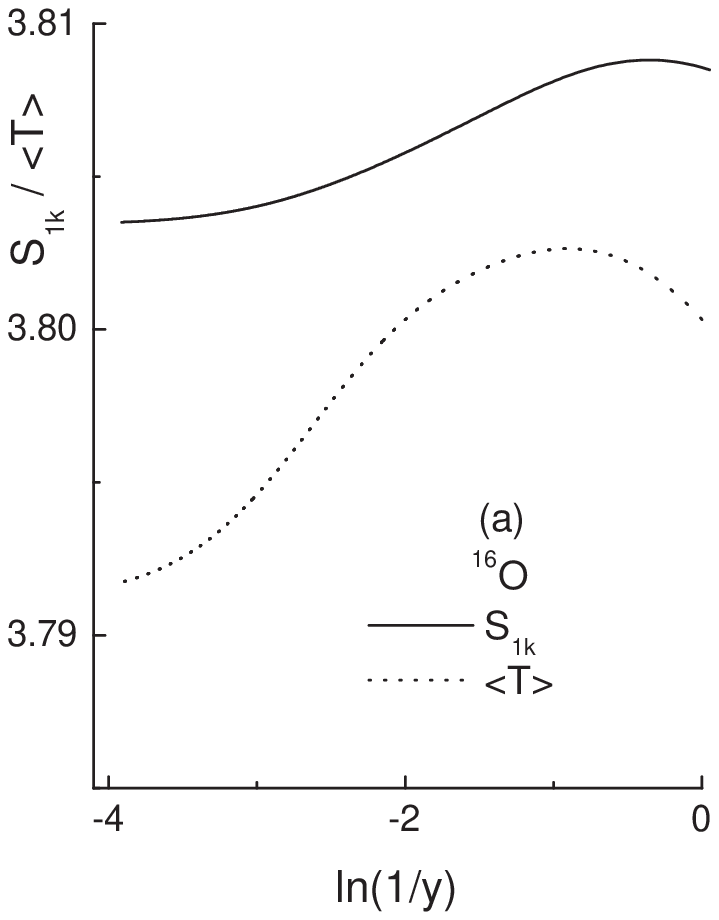}
 \hspace{0.5cm}
 \includegraphics[height=5.0cm,width=3.5cm]{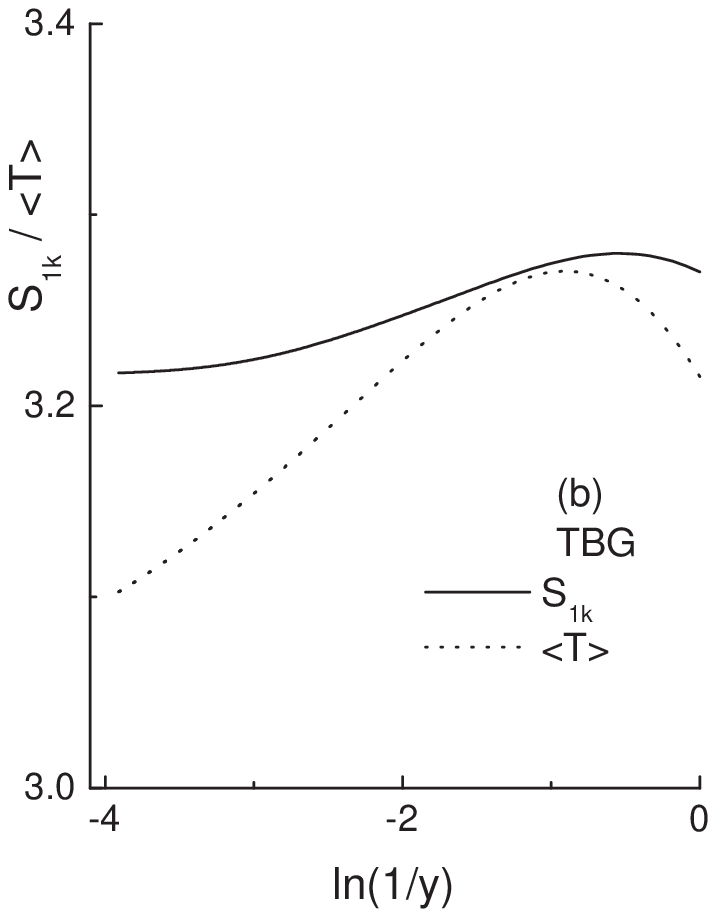}
 \caption{\label{fig:fig5} The mean kinetic energy $\langle T \rangle$ (in $\hbar
\omega$ units) and the Shannon information entropy $S_{1k}$ as a
function of the correlation parameter $\ln{(\frac{1}{y})}$, (a) in
nuclei $^{16}$O and (b) in a trapped Bose gas (TBG).}
\end{figure}

It is interesting to observe the correlation of the rms radii
$\sqrt{\langle r^2 \rangle}$ with $S_r$ as well as the
corresponding behavior of the mean kinetic energy $\langle T
\rangle$ with $S_k$, as functions of the strength of SRC
$\ln{(\frac{1}{y})}$ for the $^{16}\textrm{O}$ nucleus and trapped
Bose gas. This is done in Fig. \ref{fig:fig4} for $\sqrt{\langle
r^2 \rangle}$ and Fig. \ref{fig:fig5} for $\langle T \rangle$
after apllying the suitable rescaling. The corresponding curves
are similar for nuclei and trapped Bose gas.

\begin{figure}[h]
 \centering
 \includegraphics[height=5.0cm,width=3.5cm]{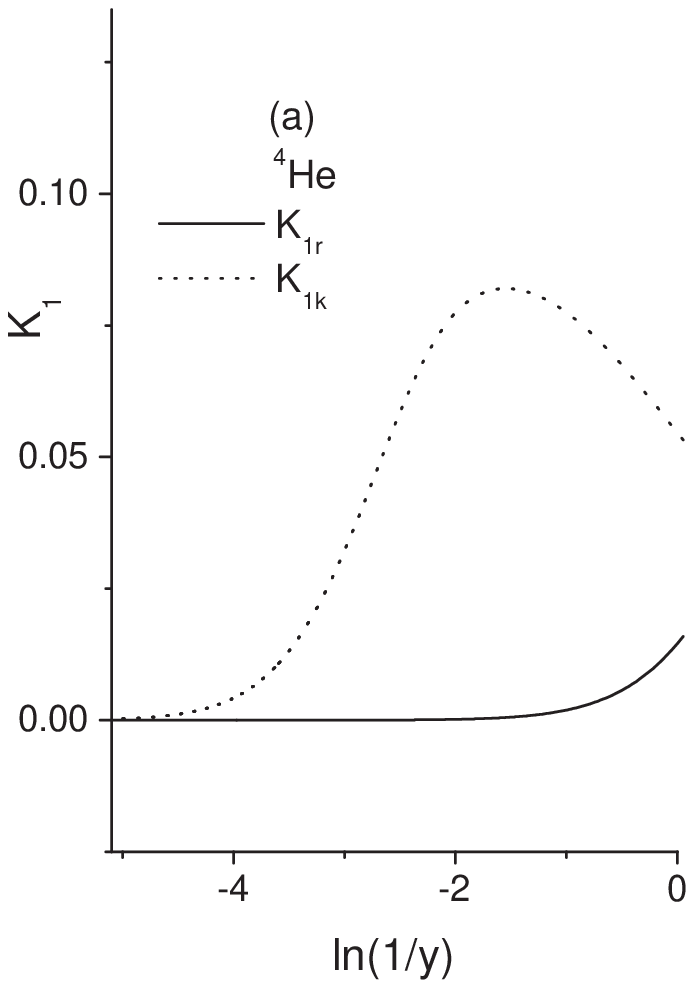}
 \hspace{0.5cm}
 \includegraphics[height=5.0cm,width=3.5cm]{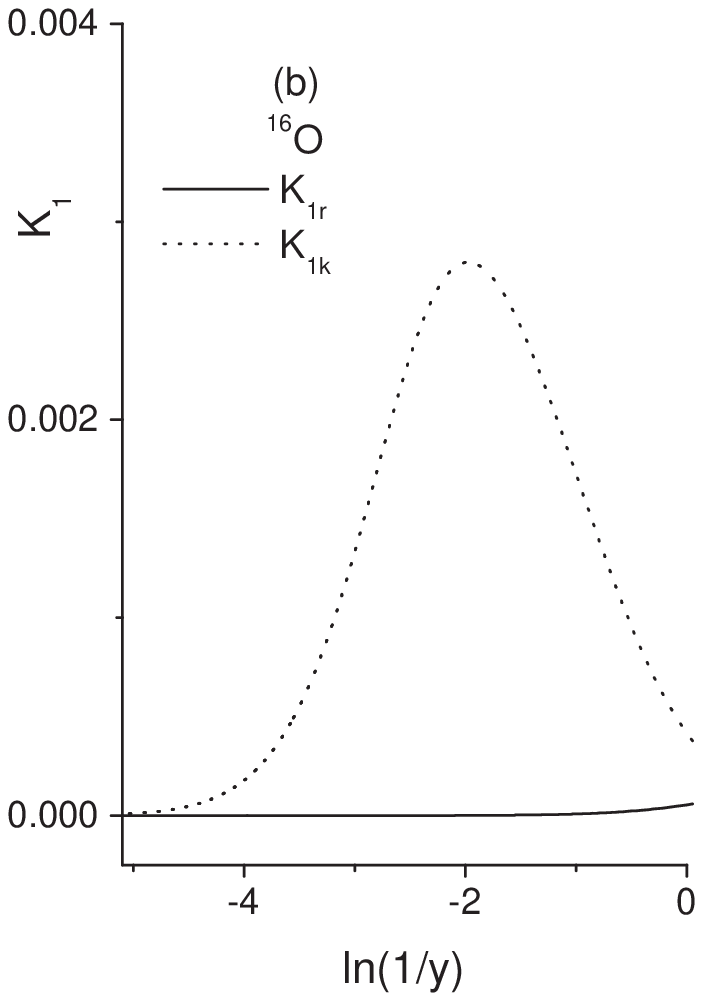}
 \hspace{0.5cm}
 \includegraphics[height=5.0cm,width=3.5cm]{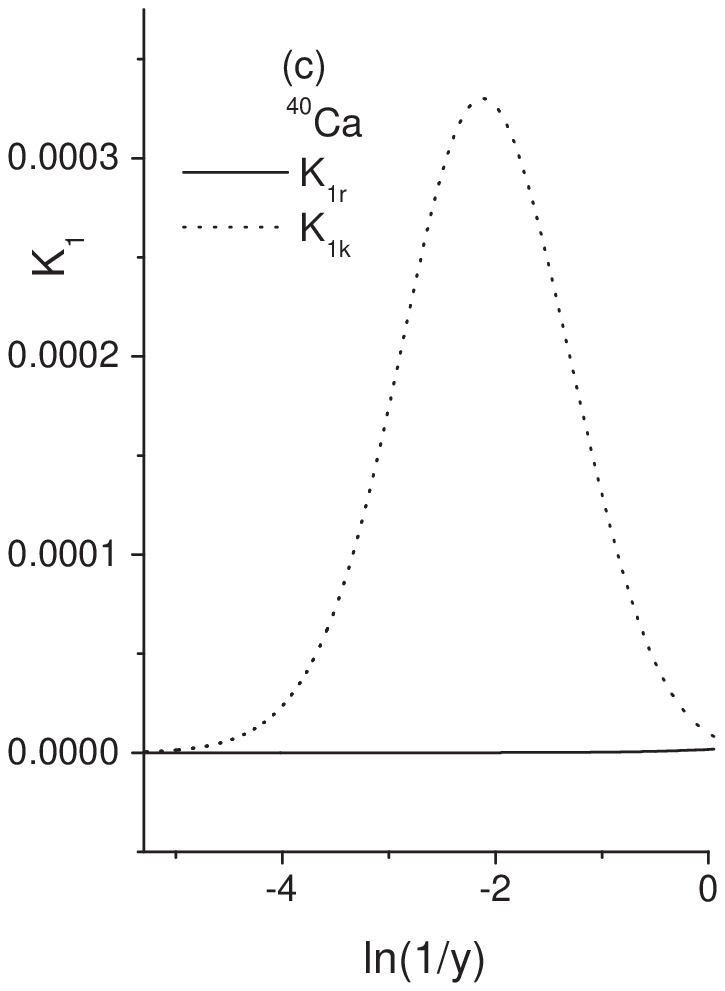}
 \caption{\label{fig:fig6} The one- body Kullback-Leibler relative entropy both in
coordinate- and momentum-space, in nuclei (a) $^{4}$He, (b)
$^{16}$O and (c)  $^{40}$Ca.}
\end{figure}

\begin{figure}[h]
 \centering
 \includegraphics[height=5.0cm,width=3.5cm]{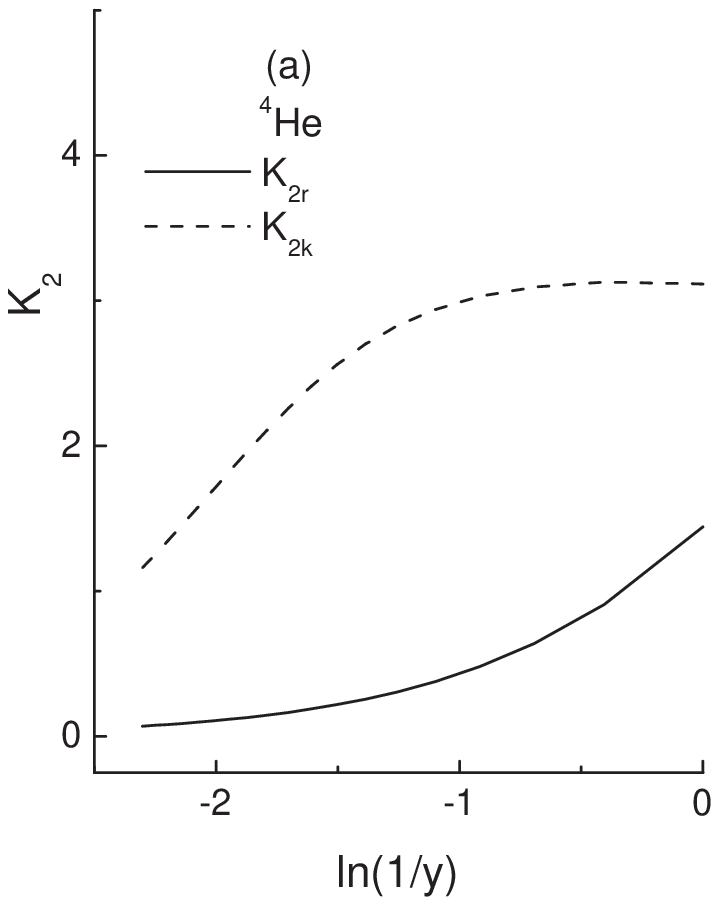}
 \hspace{0.5cm}
 \includegraphics[height=5.0cm,width=3.5cm]{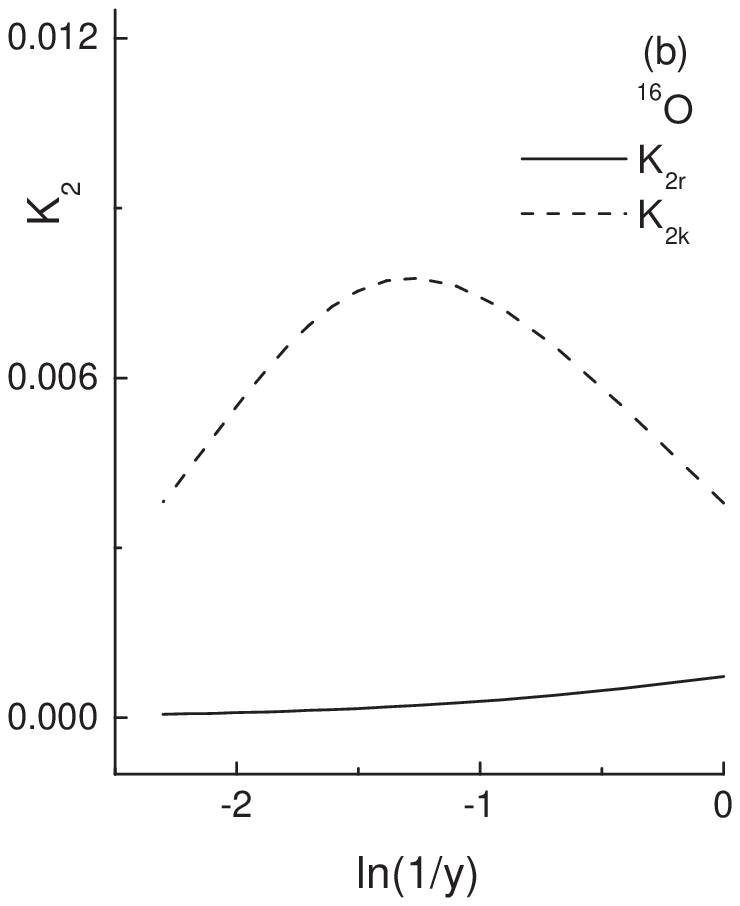}
 \hspace{0.5cm}
 \includegraphics[height=5.0cm,width=3.5cm]{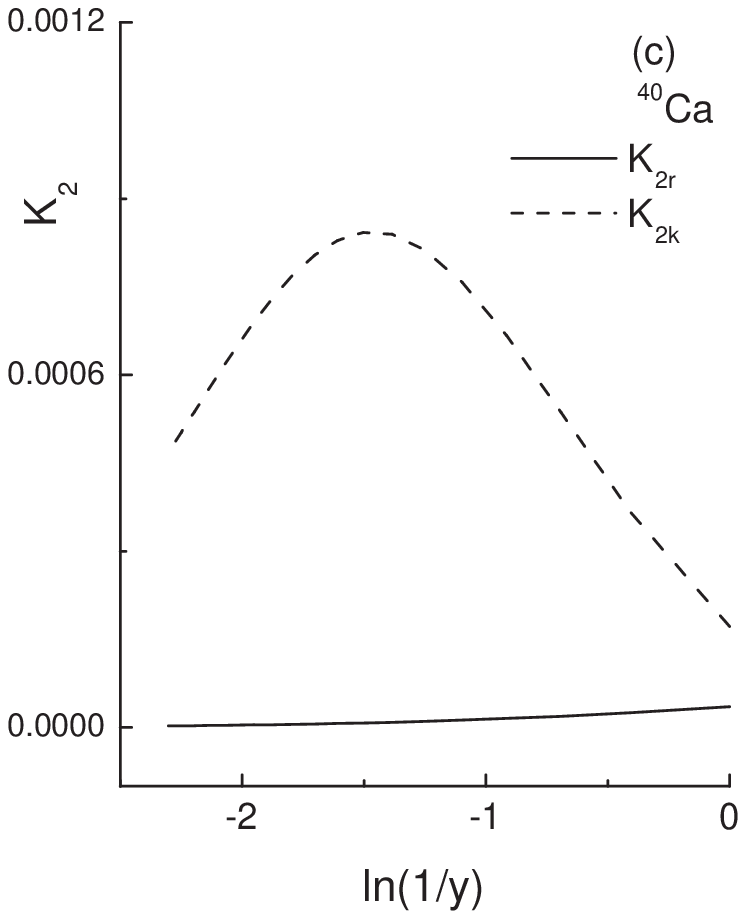}
 \caption{\label{fig:fig7} The two-body Kullback-Leibler relative entropy both in
coordinate- and momentum-space, in nuclei (a) $^{4}$He, (b)
$^{16}$O and c) $^{40}$Ca.}
\end{figure}

A well-known concept in information theory is the distance between
the probability distributions $\rho_i^{(1)}$ and $\rho^{(2)}$, in
our case the correlated and the uncorrelated distributions
respectively. A measure of distance is the Kullback-Leibler
relative entropy $K$ defined previously. The correlated and
uncorrelated cases are compared for the one-body case $(K_1)$ in
Fig. \ref{fig:fig6} and the the two-body case $(K_2)$ in Fig.
\ref{fig:fig7} for nuclei $(^4\textrm{He}, ^{16}\textrm{O},
^{40}\textrm{Ca})$ and trapped Bose gas, decomposing in position-
and momentum-spaces according to
(\ref{eq:equ17})-(\ref{eq:equ20}). It is seen that $K_{1r}$,
$K_{2r}$ increase as the strength of SRC increases, while
$K_{1k}$, $K_{2k}$ have a maximum at a certain value of
$\ln{(\frac{1}{y})}$ depending on the system under consideration.

\clearpage
\newpage

\begin{figure}[h]
 \centering
 \includegraphics[height=5.0cm,width=3.5cm]{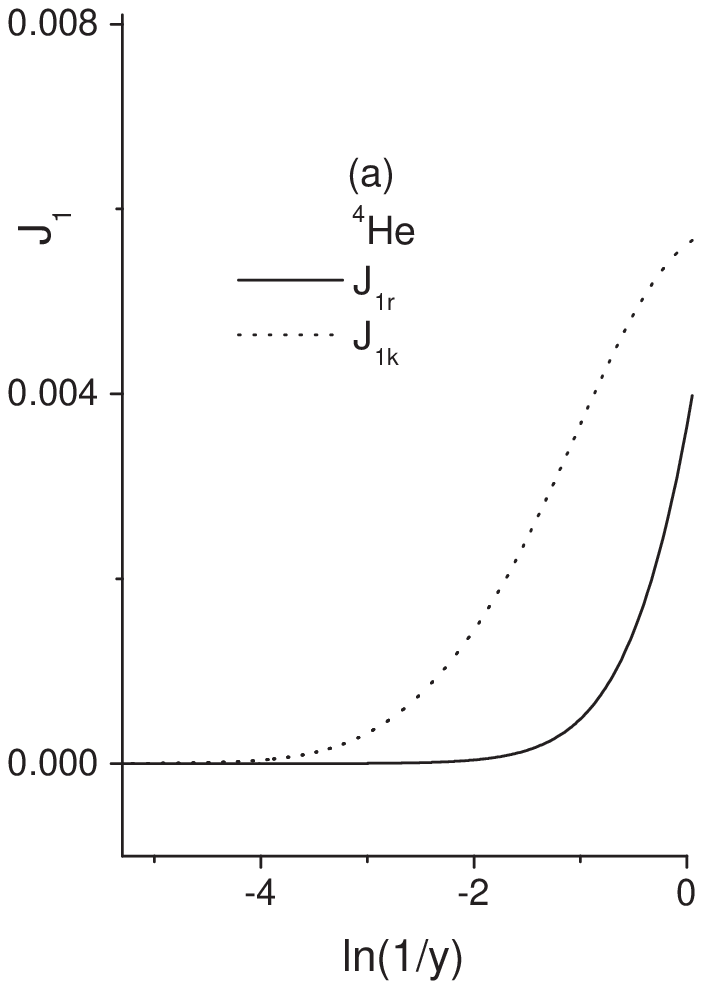}
 \hspace{0.5cm}
 \includegraphics[height=5.0cm,width=3.5cm]{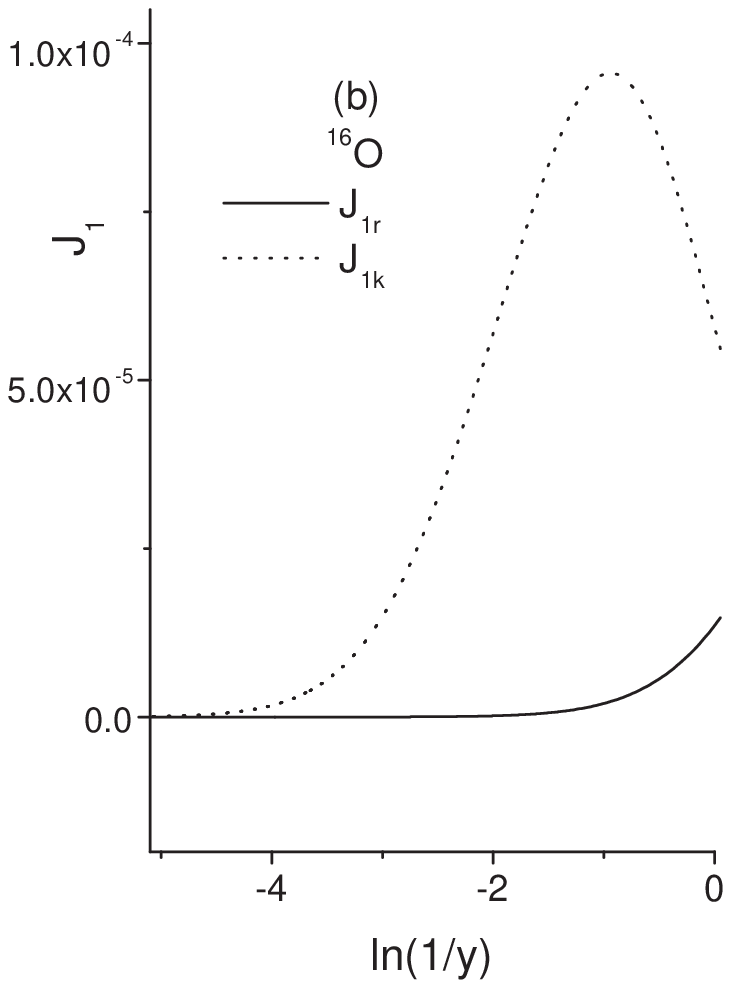}
 \hspace{0.5cm}
 \includegraphics[height=5.0cm,width=3.5cm]{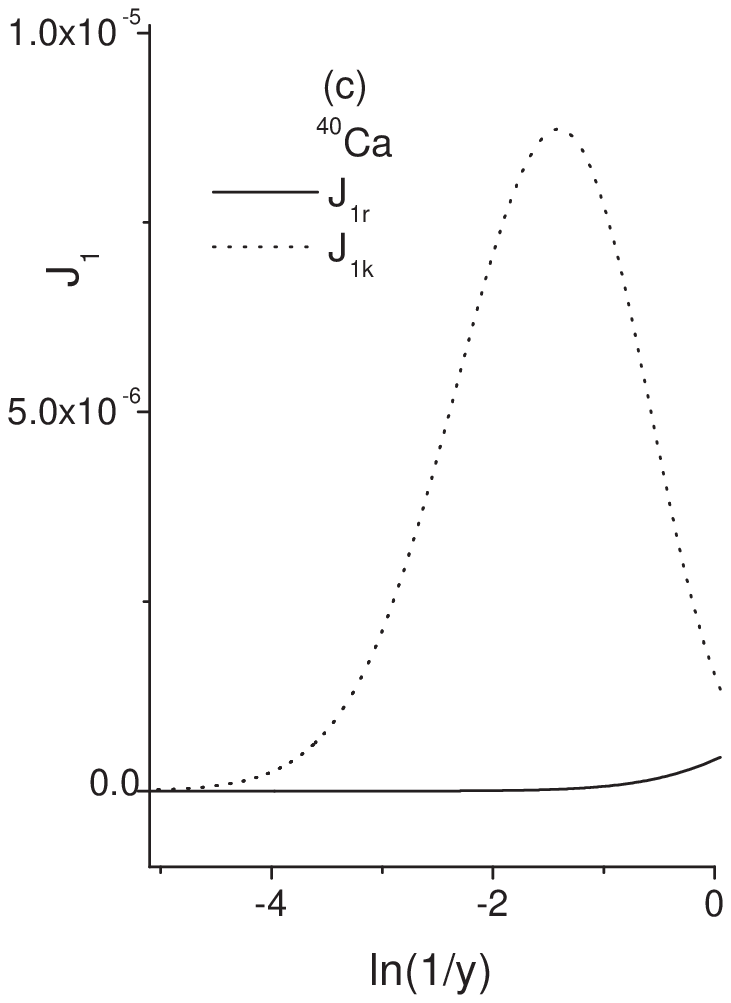}
 \caption{\label{fig:fig8} The one-body Jensen-Shannon divergence entropy both in coordinate-
and momentum-space, in nuclei (a) $^{4}$He, (b) $^{16}$O and (c)
$^{40}$Ca.}
\end{figure}

Calculations are also carried out for the Jensen-Shannon
divergence for one-body density distribution ($J_1$ entropy) as
function of $\ln{(\frac{1}{y})}$ for nuclei and trapped Bose gas,
decomposed in position- and momentum- spaces (Fig.
\ref{fig:fig8}). We observe again that $J_1$ increases with the
strength of SRC in position-space, while in most cases in
momentum-space there is a maximum for a certain value of
$\ln{(\frac{1}{y})}$. It is verified that $0<J<\ln{2}$ as expected
theoretically.\cite{Majtey}

It is noted that the dependence of the various kinds of
information entropy on the correlation parameter
$\ln{(\frac{1}{y})}$ is studied up to the value
$\ln{(\frac{1}{y})}=0$ $(y=1)$, which is already unrealistic
corresponding to strong SRC. In addition, lowest order
approximation does not work well beyond that value. In this case
three-body terms should be included but this prospect is out of
the scope of the present work.

\begin{figure}[h]
 \centering
 \includegraphics[height=5.0cm,width=3.5cm]{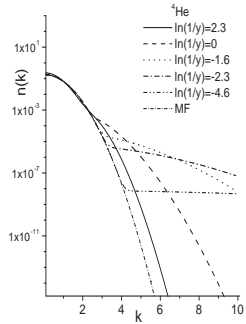}
 \caption{\label{fig:fig9} The momentum distribution $n(k)$ of $^4$He for various values
 of the correlation parameter $\ln{(\frac{1}{y})}$. The case MF (mean field)
 corresponds to the uncorrelated case $(y\rightarrow \infty).$}
\end{figure}

For very strong SRC the momentum distribution $n(k)$ exhibits a
similar behavior with the mean field $(y\rightarrow \infty)$. This
is illustrated in Fig. \ref{fig:fig9}, where we present $n(k)$ for
various values of $\ln{(\frac{1}{y})}$. It is seen that for small
and large SRC the tail of $n(k)$ disappears. That is why for small
and large SRC the relative entropy ($K_{1k}$ and $J_{1k}$) is
small, while in between shows a maximum (Fig. \ref{fig:fig6},
\ref{fig:fig8}). A similar trend of $n(\textbf{k}_1,\textbf{k}_2)$
for large SRC explains also the maximum of the relative entropy
$K_{2k}$ in Fig. \ref{fig:fig7}.

\section{Conclusions and final comment}\label{sec:9}

Our main conclusions are the following
\begin{itemize}
\item[(i)] Increasing the SRC (i.e. the parameter
$\ln{(\frac{1}{y})}$) the information entropies $S$, $O$, $K$ and
$J$ increase. A comparison leads to the conclusion that the
correlated systems have larger values of entropies than the
uncorrelated ones.

\item[(ii)] There is a similar behavior of the entropies as
functions of correlations for both systems (nuclei and trapped
Bose gas) although they obey different statistics (fermions and
bosons).

\item[(iii)] There is a correlation of $\sqrt{\langle r^2
\rangle}$ with $S_r$ and $\langle T \rangle$ with $S_k$ in the
sense that they have the same behavior as a function of the
correlation parameter $\ln{(\frac{1}{y})}$. These results can lead
us to relate the theoretical quantities $S_r$ and $S_k$ with
experimental ones like charge form factor, charge density
distribution, and momentum distribution, radii, etc. A recent
paper addressed in that problem.\cite{Massen-05}

\item[(iv)] The relations $S_2=2 S_1$ and $O_2=O_1^2$ hold exactly
for the uncorrelated densities in trapped Bose gas while the above
relations are almost exact for the uncorrelated densities and in
the case of correlated densities both in nuclei and trapped Bose
gas. In previous work we proposed the universal relation
$S_1=S_r+S_k=a+b\,\ln{N}$ where $N$ is the number of particles of
the system either fermionic (nucleus, atom, atomic cluster) or
bosonic (correlated atoms in a trap). Thus in our case
\[
  S_2=2(a+b\,\ln{N})
\]
For 3-body distributions
$\rho(\textbf{r}_1,\textbf{r}_2,\textbf{r}_3)$ and
$n(\textbf{k}_1,\textbf{k}_2,\textbf{k}_3)$
\[
  S_3=3\,(a+b\,\ln{N})
\]
and generalizing for the $N$-body distributions
$\rho(\textbf{r}_1,\textbf{r}_2,\ldots,\textbf{r}_N)$ and
$n(\textbf{k}_1,\textbf{k}_2,\ldots,\textbf{k}_N)$
\[
  S_N=N\,(a+b\,\ln{N})
\]
This is exact for the uncorrelated trapped Bose gas, almost exact
in correlated nuclei $(N=1,2)$ and it is conjectured that it holds
approximately for correlated systems (which has still to be proved
for $N\geq 3$).

\item[(v)] The entropic uncertainty relation (EUR) is
\[
  S=S_r+S_k\geq 6.434
\]
It is well-known that the lower bound is attained for a Gaussian
distribution (i.e. the case of $^4\textrm{He}$ uncorrelated). In
all cases studied in the present work EUR is verified.

A final comment seems appropriate. In general, the calculation of
$\rho({\bf r}_1,{\bf r}_2)$ and $n({\bf k}_1,{\bf k}_2)$ is a
problem very hard to be solved, especially in the case of nuclei,
in the framework of short range correlations. Just a few works are
addressed in that
problem.\cite{Bohigas,Dimitrova-00,Papakonstantinou} In the
present work we tried to treat the problem in an approximate but
self-consistent way in the sense that the calculations of
$\rho({\bf r}_1,{\bf r}_2)$ and $n({\bf k}_1,{\bf k}_2)$ are based
in the same $\rho({\bf r}_1,{\bf r}_2;{\bf r}_1',{\bf r}_2')$,
which is the generating function of the above quantities. As a
consequence the information entropy $S_2=S_{2r}+S_{2k}$ is derived
also in a self-consistent way and there is a direct link between
$S_{2r}$ and $S_{2k}$, as well as the other kinds of information
entropies which are studied in the present work.
\end{itemize}

\section*{Acknowledgments}

The work of Ch.~C. Moustakidis was supported by the Greek State
Grants Foundation (IKY) under contract (515/2005) while the work
of K.~Ch. Chatzisavvas by Herakleitos Research Scolarships
(21866). One of the authors (Ch.~C.~M.) would like to thank Prof.
Vergados for his hospitality in the University of Ioannina where
the earlier part of this work was performed.


\clearpage
\newpage

\begin{thebibliography}{99}

\bibitem{Ohya}
 M.~Ohya,  and D.~Petz, \emph{Quantum Entropy and Its Use} (
 Springer-Verlag, Berlin; New York, 1993).

\bibitem{Bialynicki}
 I.~Bialynicki-Birula, and J.~Mycielski, \emph{Commun. Math. Phys}
 \textbf{44} (1975) 129.

\bibitem{Panos1}
 C.~P.~Panos, and S.~E.~Massen, \emph{Int. J. Mod. Phys.}
 \textbf{E6} (1997) 497 .

\bibitem{Massen1}
 S.~E.~Massen, Ch.~C.~Moustakidis, and C.~P.~Panos, \emph{Phys. Let.}
 \textbf{A64} (2002) 131.

\bibitem{Massen2}
S.~E.~Massen, and C.~P.~Panos, \emph{Phys. Lett.} \textbf{A246}
(1998) 530.

\bibitem{Massen3}
 S.~E.~Massen, and C.~P.~Panos, \emph{Phys. Lett.} \textbf{A280} (2001)
 65.

\bibitem{Garde1}
 S.~R.~Gadre, S.~B.~Sears, S.~J.~Chakravorty, and R.~D.~Bendale,
 \emph{Phys. Rev.} \textbf{A32} (1985) 2602.

\bibitem{Garde2}
 S.~R.~Gadre, and R.~D.~Bendale,
 \emph{Phys. Rev.} \textbf{A36} (1987) 1932.

\bibitem{Ghosh}
 S.~K.~Ghosh, M.~Berkowitz, and R.~G.~Parr, \emph{Proc. Natl. Acad.
 Sc. USA} \textbf{81} (1984) 8028.

\bibitem{Lalazissis}
 G.~A.~Lalazissis, S.~E.~Massen, C.~P.~Panos, and S.~S. Dimitrova,
 \emph{Int. J. Mod. Phys.} \textbf{E7} (1998) 485.

\bibitem{Moustakidis}
 Ch.~C.~Moustakidis, S.~E.~Massen, C.~P.~Panos, M.~E.~Grypeos,
 and A.~N.~Antonov, \emph{Phys. Rev.} \textbf{64} (2001) 014314.

\bibitem{Panos2}
 C.~P.~Panos, S.~E.~Massen, and C.~G.~Koutroulos, \emph{Phys. Rev.}
 \textbf{63} (2001) 064307.

\bibitem{Panos3}
 C.~P.~Panos, \emph{Phys. Lett.} \textbf{A289} (2001) 287.

\bibitem{Massen4}
 S.~E.~Massen, \emph{Phys. Rev.} \textbf{C67} (2003) 014314.

\bibitem{Moustakidis1} Ch.C. Moustakidis, and S.E. Massen, \emph{Phys. Rev.} {\bf
B71} (2003) 045102.

\bibitem{Massen5}
 S.~E.~Massen, Ch.~C.~Moustakidis, and C.~P.~Panos, \emph{Focus on Boson Research}
 (Nova Publishers, editor A.~V.~Ling) In press.

\bibitem{Chatzisavvas}
 K.~Ch.~Chatzisavvas, and C.~P.~Panos, to be published in \emph{Int. J. Mod Phys. E}
 (2005).

\bibitem{Fabrocini}
 A.~Fabrocini, and A.~Polls, \emph{Phys. Rev.}  \textbf{A60} (1999) 2319.

\bibitem{Moustakidis2}
 Ch.~C. Moustakidis, and S.~E.~Massen, \emph{Phys. Rev.} {\bf
 A65} (2002) 063613.

\bibitem{Jastrow}
 R.~Jastrow, \emph{Phys. Rev.} {\bf 98} (1955) 1497.

\bibitem{Onicescu}
 O.~Onicescu, \emph{R. Acad. Sci. Paris} {\bf A263} (1996) 25.

\bibitem{Kullback}
 S.~Kullback, \emph{Statistics and Information Theory}, Wiley, New
 York, (1959).

\bibitem{Rao}
 C.~Rao, \emph{Differential Geometry in Statistical Interference},
 IMS-Lectures Notes, \textbf{10} (1987) 217.

\bibitem{Lin}
 J.~Lin, \emph{IEEE Trans. Inf. Theory} \textbf{37} 1 (1991) 145.

\bibitem{Majtey}
 A.~Majtey, P.~W.~Lamberti, M.~T.~Martin, and A.~Plastino,
 quant-ph/0408082.

\bibitem{Lepadatu}
 C.~Lepadatu, and E.~Nitulescu, \emph{Acta Chim. Slov.}
 \textbf{50} (2003) 539.

\bibitem{Lowdin}
 P.~O.~Lowdin, \emph{Phys. Rev} \textbf{97} (1955) 1474.

\bibitem{Amovilli}
 C.~Amovilli, N.~H.~March, \emph{Phys. Rev.} \textbf{A69} (2004) 054302.

\bibitem{Cover}
 T.~M.~Cover, and J.~A.~Thomas, \emph{Elements of Information Theory},
 (Wiley-Interscience, New York 1991).

\bibitem{Massen6}
 S.~E.~Massen, and Ch.~Moustakidis, \emph{Phys. Rev. }\textbf{C60} (1999)
 024005; Ch.~Moustakidis, and S.~E.~Massen, Phys. Rev. \textbf{C62} (2000)
 034318.

\bibitem{Stringari} S.~Stringari, M.~Traini, O.~Bohigas,
\emph{Nucl. Phys.} \textbf{A516} (1990) 33.

\bibitem{Gaidarov}
 M.~K.~Gaidarov, A.~N.~Antonov, G.~S.~Anagnostatos, S.~E.~Massen,
 M.~V.~Stoitsov, P.~E.~Hodgson, \emph{Phys. Rev.} \textbf{C52} (1995)
 3026.

\bibitem{Massen-05} S.E.~Massen, V.P.~Psonis, A.N.~Antonov, e-print
nucl-th/0502047.

\bibitem{Bohigas}
 O.~Bohigas, and S.~Stringari, \emph{Phys. Lett} \textbf{B95} (1980) 9;
  M.~Dal.~Ri, S.~Stringari, and O.~Bohigas, \emph{Nucl. Phys.}
\textbf{A376} (1982) 81.

\bibitem{Dimitrova-00}S.S.~Dimitrova, D.N.~Kadrev, A.N.~Antonov,
and M.V.~Stoitsov, \emph{Eur. Phy. J.} \textbf{A7} (2000) 335.

\bibitem{Papakonstantinou}
P.~Papakonstantinou, E.~Mavrommatis, and T.~S.~Kosmas,\emph{ Nucl.
Phys.} \textbf{A713} (2003) 81.

\end{thebibliography}
\end{document}